\documentclass[11pt]{article}

\usepackage{multicol}
\usepackage{geometry}
\usepackage{dcolumn}
\usepackage{amsmath}
\usepackage{threeparttable}

\usepackage{lmodern}
\usepackage{graphicx} 
\usepackage{enumitem}
\usepackage{lscape}
\usepackage{setspace}
\usepackage{booktabs}
\usepackage{amssymb}

\usepackage[gen]{eurosym}
\usepackage{pdflscape}      
\usepackage{booktabs}       
\usepackage{threeparttable}
\usepackage{adjustbox}
\usepackage{textcomp}       
\usepackage{siunitx}        

\usepackage{newtxtext,newtxmath}  

\usepackage[authoryear,round]{natbib}   

\usepackage{xcolor}
\usepackage{hyperref}
\usepackage{hyperref}

\definecolor{mahogany}{RGB}{192,64,0}

\hypersetup{
  colorlinks=true,      
  linkcolor=mahogany,        
  urlcolor=mahogany,         
  citecolor=mahogany,        
  filecolor=mahogany,        
  pdfusetitle=true,     
  urlbordercolor=mahogany,   
  linkbordercolor=mahogany,  
}

\begin{document}



\title{Spatial Competition on Psychological Pricing Strategies --
Preliminary Evidence from an Online Marketplace}

\author{%
  Magdalena Schindl\thanks{Institute of Data and Knowledge Engineering, Johannes Kepler University Linz, 4040 Linz, Austria.}%
  \and%
  Felix Reichel\thanks{Department of Economics, Johannes Kepler University Linz, 4040 Linz, Austria. \newline \textit{Corresponding author}: \texttt{felix.reichel@jku.at}}%
}

\date{\today}

\maketitle
\thispagestyle{empty}

\begin{abstract}
This paper investigates whether spatial proximity shapes psychological-pricing choices on Austria’s C2C marketplace \textit{willhaben}. Two web-scraped snapshots of 826 \textbf{Woom Bike} listings a standardised product sold on the platform reveal that sellers near direct competitors are more likely to adopt 9-, 90-, or 99-ending prices, who also use such pricing strategy unconditional on product characteristics or underlying spatiotemporal differences. Such strategy is associated with an average premium of approximately cet. par. 3.4 \%. Information asymmetry persists: buyer trust hinges on signals such as the “Trusted Seller” badge, and missing data on the “PayLivery” feature. Lacking final transaction prices limits inference.

\end{abstract}

\vspace{0.6em}

\vspace{0.3em}

\newpage

\newpage

\begin{center}
    \Large\textbf{Disclosure of Conflicts Statement}
\end{center}

\vspace{1cm}

\noindent
This document confirms that there are no conflicts of interest associated with this work. All contributors have reviewed and affirmed that no personal, financial, or professional conflicts exist that could influence the results or conclusions presented.

\vspace{1cm}

\newpage

\tableofcontents

\newpage

\newpage

\newpage

\section{Introduction}

According to \citet{blockchain}, online marketplaces facilitate the purchase and sale of products and services, as well as the exchange of money and data between users and the platform. Their large product selection, low costs, the convenience of shopping without physical constraints, and their technical possibilities have driven rapid growth (\citet{blockchain}). Online marketplaces also operate in the consumer-to-consumer (C2C) sector, providing a broad user group with a venue for trading used products. This article therefore focuses on \citet{willhaben}, Austria’s leading C2C marketplace, as identified by \citet{grund}.
The empirical analysis in this paper centers on the listings for \textbf{Woom Bikes}, a standardised product sold on Willhaben. Through web scraping of two snapshots, we created a dataset of roughly 900 observations, focusing on mid- to high-price segment bicycles, which we claim are characterised by price stability and uniformity. This analysis examines ad-listing prices through predictive models that use Willhaben product-listing attributes and the spatial distribution of one constructed attribute—psychological pricing (e.g., 99-ending prices).
 Our paper thus tackles the following \textbf{research questions}:
\begin{enumerate}
    \item How does information asymmetry between buyers and sellers affect overall market efficiency and transactions?
    \item How much of the variance in Woom bike prices can we explain using a multiple linear regression (MLR) model based on observed product characteristics obtained through web scraping (e.g., size, color, condition, and engineered/encoded variables such as psychological pricing indicators)?
    \item Is there spatial clustering in psychological pricing strategies present?
\end{enumerate}


\newpage

\section{Theoretical Part: Online Marketplaces -- The Case of \textit{willhaben.at}, a Primarily C2C Platform in Austria}

Willhaben Internet Service GmbH~\&~Co~KG was founded in Austria in~2006. \textit{Willhaben} is a virtual marketplace focusing on exchanges between $n$ buyers and $m$ sellers. As of October~2021, \textit{willhaben.at} recorded 82\,127\,327 visits and 1\,510\,387\,470 page views, 93\% of which originated in Austria (\citet{oewa}).

\textit{Willhaben} primarily serves as a classifieds platform and does not employ an explicit allocation mechanism such as an auction implementation. Users simply list items for sale, and interested buyers contact sellers directly to obtain further information and, where appropriate, negotiate prices. With the help of “PayLivery”, \textit{Willhaben} offers, for a small fee, a service that helps secure transactions. Hence, the marketplace functions more as a direct exchange or matching mechanism than as an auction-based system. On the platform, both buyers and sellers may be private individuals or businesses.

The following companies or marketplaces can be named as main competitors: eBay, Vinted, kleinanzeigen.de, offline second-hand shops and flea markets, as well as private garage sales. The products offered on \textit{willhaben.at} are mostly used items and can be either standardised goods or handmade products. Obersteiner et~al.\ (2023) also examine \textit{Willhaben} and note that the platform provides a marketplace for used cars and for apartments and houses for rent or purchase; these product areas are not considered in our work.

According to \citet{werbege}, regional comparison platforms such as Check24 and Geizhals likewise compete in the e-commerce landscape, offering companies new opportunities for information and purchasing processes. These platforms provide insight into regional search trends and thus give valuable guidance for selecting suitable sales channels and understanding geographical and temporal differences in consumer behaviour (\citet{werbege}).

Hu et~al.\ (2023) state that the disclosure of product information on online platforms can yield both positive and negative effects. The authors distinguish between a \emph{seller’s market}—where buyers outnumber sellers—and a \emph{buyer’s market}—where sellers outnumber buyers. In a seller’s market, revealing product information benefits the platform because it helps buyers evaluate individual products and can raise their willingness-to-pay (WTP). By contrast, in a buyer’s market, withholding certain information may be advantageous: it reduces variation in product evaluations, increases the likelihood that buyers possess sufficient WTP, and can ultimately boost overall sales (\citet{prodandmarketin}).
\noindent According to \citet{willhaben}, sellers on the platform can provide the following details for the bicycle model investigated in our empirical section:
\begin{multicols}{2}
\begin{itemize}
  \item title
  \item point of sale; PayLivery option
  \item photos/images
  \item price
  \item description
  \item type of brake
  \item wheel size
  \item frame size (indicated by version number)
  \item colour
  \item condition (used, new, almost like new)
  \item transfer
  \item seller: name, location, private or business
\end{itemize}
\end{multicols}

While these attributes are available to sellers, our empirical analysis focuses only on a small subset relevant to bicycles. Product photographs, which could be valuable for marketing, were excluded from the dataset because they would require post-scraping classification and therefore were not scraped in the first place.
To register as a buyer or private seller, all you need is an e-mail address and a device with internet access; otherwise there are no restrictions. During registration, the “Terms and Conditions” and a CAPTCHA must also be confirmed.
Nevertheless, \textit{willhaben} now offers many functions that distinguish itself from the competitor marketplaces mentioned earlier. Furthermore, the platform's regional focus simplifies transactions and logistics and often allows buyers and sellers to meet in person, pick up the item, and view it before buying. Sellers can choose how the item is to be sold—pickup, shipping, PayLivery, or a combination of these options. The PayLivery option means that the payment process and the shipping label are handled by \textit{willhaben}, offering a secure transaction for both parties; the platform charges a service fee for this. In addition, users can list items free of charge, which attracts a large number of private sellers, making it a buyer's market for most popular used goods. One of \textit{willhaben}’s revenue models is to highlight listings for a fee.
Because the platform offers a wide range of product categories—from real estate and vehicles to electronics, fashion, sporting goods, books, and even jobs—\textit{willhaben} serves many buyers and sellers. The platform also allows users to remain anonymous in their transactions, which can be attractive for those who value privacy. Furthermore, \textit{willhaben} offers a “Trusted Seller” programme that highlights sellers with positive reviews and thus increases buyer confidence. Together, these services provide many advantages for buyers and sellers and underpin \textit{willhaben}’s strong market leadership position, as documented by the statistics by \citet{oewa}.
Although product information is very important, \citet{prodandmarketin} point out that other factors also influence buyers’ decisions (\citet{c2c}). Moriuchi and Takahashi~(2022) investigated consumer behaviour in the consumer-to-consumer (C2C) e-commerce marketplace in Japan, focusing on the roles of value, trust, and engagement in buyer satisfaction.
Their study extends the Means–End Chain (MEC) theory by integrating trust, engagement, and satisfaction as critical factors. Product quality affects the value buyers perceive, the shopping pleasure they experience, and the emotional value derived; such emotional connection is crucial for the overall brand experience.

\citet{c2c} further found that functional value exerts a stronger influence on trust—both in the intermediary and in the seller—than emotional value. They conclude that buyers primarily seek products that meet their needs, although emotional engagement can also enhance satisfaction. Trust in the intermediary (e.g., a platform like \textit{willhaben}) is crucial for customer loyalty, yet it has no direct influence on satisfaction. By contrast, trust in the seller directly affects satisfaction, suggesting that active interaction between buyers and sellers leads to higher satisfaction (\citet{c2c}).

Market liquidity refers to the ease with which assets can be bought or sold in a marketplace without significantly affecting their price. In lateral‐exchange marketplaces—such as those enabled by platforms like Airbnb or eBay—liquidity depends on how efficiently the platform matches supply and demand. The concept of \emph{liquid ownership}, meaning temporary access to goods rather than full ownership, is central to understanding these marketplaces. As \citet{platf} note, platforms like \textit{willhaben} can benefit from distinguishing between marketplace models such as access‐based consumption (e.g.\ renting) and collaborative consumption (e.g.\ sharing‐economy services). For \textit{willhaben}, improving liquidity means ensuring that users can readily access and exchange goods, thus enhancing the platform’s overall efficiency and appeal. Recognising this market segmentation can help \textit{willhaben} position itself more effectively, optimise its features, and compete in the ever evolving digital‐marketplace landscape.

Newer technologies could further improve online second‐hand marketplaces. According to Kadir et~al.\ (2023), the second‐hand sector—especially for high‐value items such as luxury goods—faces persistent challenges of trust, transparency, and product authenticity. The authors thus propose blockchain technology to address these issues: its secure and immutable ledger can ensure ownership traceability and product authenticity. For example, the OWNTRAD platform employs blockchain to enhance transparency and trust in vintage e-commerce by allowing users to trace ownership transactions and verify the quality and legality of listed items. Kadir et~al.\ (2023) report that such systems not only increase customer satisfaction but also improve profitability by reinforcing consumer trust.

\noindent \textit{Willhaben} could adopt similar blockchain‐based solutions to mitigate trust and transparency concerns on its second‐hand marketplace, building on the insights of (Kadir et~al.\ (2023)).

\newpage
\subsection{Advantages and Disadvantages on the \textit{willhaben} Marketplace}

\begin{table}[h]
\centering
\begin{tabular}{p{0.22\textwidth} p{0.38\textwidth} p{0.38\textwidth}}
\toprule
\textbf{} & \textbf{Advantages} & \textbf{Disadvantages} \\
\midrule
\textbf{Buyers} &
\begin{itemize}[leftmargin=*]
    \item Product range and local focus: lower shipping costs; in-person pickup.
    \item Ease of use and accessibility: registration with e-mail only.
    \item Price negotiation: direct chat or PayLivery offers.
\end{itemize}
&
\begin{itemize}[leftmargin=*]
    \item Information asymmetry: seller-provided data may be incomplete.
    \item Limited buyer protection: higher risk for costly items.
\end{itemize}
\\
\midrule
\textbf{Sellers} &
\begin{itemize}[leftmargin=*]
    \item Large user base: millions of visits; high visibility.
    \item Cost-effective selling: free listings for private users.
    \item Flexibility in pricing: set own prices and descriptions.
\end{itemize}
&
\begin{itemize}[leftmargin=*]
    \item Competition: similar items, local price pressure.
    \item Length of sales process: inquiry, negotiation, pickup.
    \item Buyer trust: proving quality and authenticity, especially for high-value goods.
\end{itemize}
\\
\bottomrule
\end{tabular}
\end{table}

\newpage

\section{Empirical Part: An Analysis of \textit{willhaben} Ads for a Single Product}

\subsection{Introduction}


For the empirical analysis of our online marketplace, we focus on a single  highly standardized product: \textbf{“Woom Bikes”}, categorized under \textit{Marketplace}~$\rightarrow$~\textit{Sport / Sports Equipment}~$\rightarrow$~\textit{Bicycles / Cycling}~$\rightarrow$~\textit{Bicycles}. Data were collected in two web-scraper executions, initially yielding roughly 3,000 observations. After removing duplicates, handling missing values, and merging with auxiliary data sets, about 900 observations remained. A further refinement—excluding price outliers beyond 1.5 interquartile ranges (IQR)—resulted in a final sample of 826 observational units. We claim that this product is a suitable unit for analysing ads in the mid- to high-price segment, which is marked by price stability and product uniformity, varying mainly in size and colour.

\subsection{Data}

In addition to the web-scraped data of Woom Bikes\footnote{Woom Bikes on a Primarily C2C Online Exchange Marketplace: \url{https://www.willhaben.at/iad/kaufen-und-verkaufen/marktplatz/fahrraeder-radsport/fahrraeder-4552?sfId=b8725e40-07af-41a5-bb6d-6d32deed8220&rows=30&isNavigation=true&keyword=woom+4}}, we utilised further data sources to augment the initial dataset. Zip codes were provided by the Austrian parcel-delivery service Post.at\footnote{Austrian Zip Codes Data: According to ISO standard: \url{https://www.post.at/en/g/c/postal-encyclopedia}}, and geocoded data were used to link these zip codes to longitude and latitude coordinates\footnote{Geocoded Zip Codes: \url{https://gist.github.com/PeterTheOne/7135a370b37406e6801d36827e0316cf}}. Additionally, shapefiles of Austria with grid sizes of 1\,km, 10\,km, and 100\,km were obtained from the European Environment Agency (EEA)\footnote{Shapefiles of Austria (Grid 1, 10, and 100 kilometres): \url{https://www.eea.europa.eu/data-and-maps/data/eea-reference-grids-2/gis-files/austria-shapefile}}. 

Currently not included in our analysis, but potentially valuable, are datasets such as the official locations of Woom Bike resellers who sell new bikes to customers, obtainable from a dealer-locator website\footnote{All Woom B2C Resellers with Zip Codes and Country: \url{https://intl-checkout.woom.com/apps/dealerlocator}}. Other useful datasets might include the official prices of new products from the bicycle manufacturer itself\footnote{Woom Official Website: \url{https://woom.com/de_AT/}} and additional Austria-specific shapefiles—such as those depicting municipal zip-code-level borders—which would be beneficial for further empirical analysis or for investigating spatial and/or spatiotemporal dependencies among observational units while controlling for underlying spatial dependencies at the municipal zip-code level.

\subsection{Descriptive Statistics}

\begin{landscape}
\begin{figure}[!htbp]
  \centering
  \includegraphics[width=1.1\linewidth]{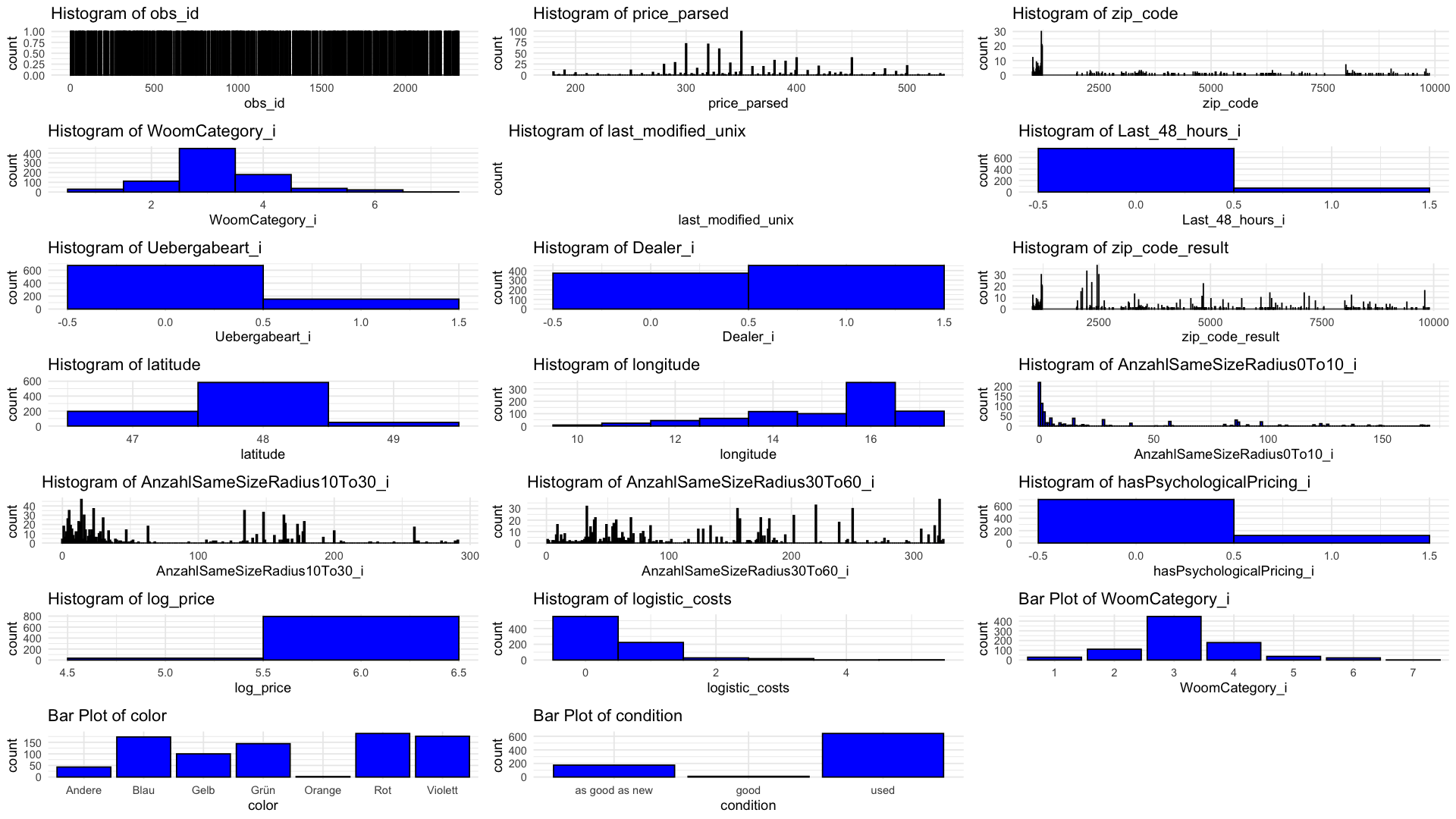}
  \caption{Histograms of key variables.\newline
           \footnotesize \textit{ Note:} Price outliers beyond 1.5\,IQR have been removed. Two snapshots; \(n = 826\).}
  \label{fig:hists}
\end{figure}
\end{landscape}

Figure~\ref{fig:hists} shows descriptive statistics in the form of histograms for all numerical variables as well as the categorical variables: size (\textit{WoomCategory}), color, and condition. Price outliers have already been removed, resulting in \( n = 826 \) observations. Several variables exhibit imbalanced distributions. For example, there are almost no listings for orange-colored bikes. Furthermore, many independent variables are imbalanced binary (0/1) or count variables. In terms of condition, most entries are either listed as \textit{as good as new} or \textit{used}, with only a small number labeled as \textit{good}. Likewise, the distribution of sizes is skewed, with most bikes falling into size categories 3 or 4. 

The histograms for zip codes and the difference count variables—based on bikes of the same size and condition—suggest some clustering, possibly reflecting city centers or urban areas, or differences in population density. Nonetheless, the spatial distribution of bike listings appears fairly balanced overall.

\begin{landscape}
\begin{figure}[!htbp]
  \centering
  \includegraphics[width=1.1\linewidth]{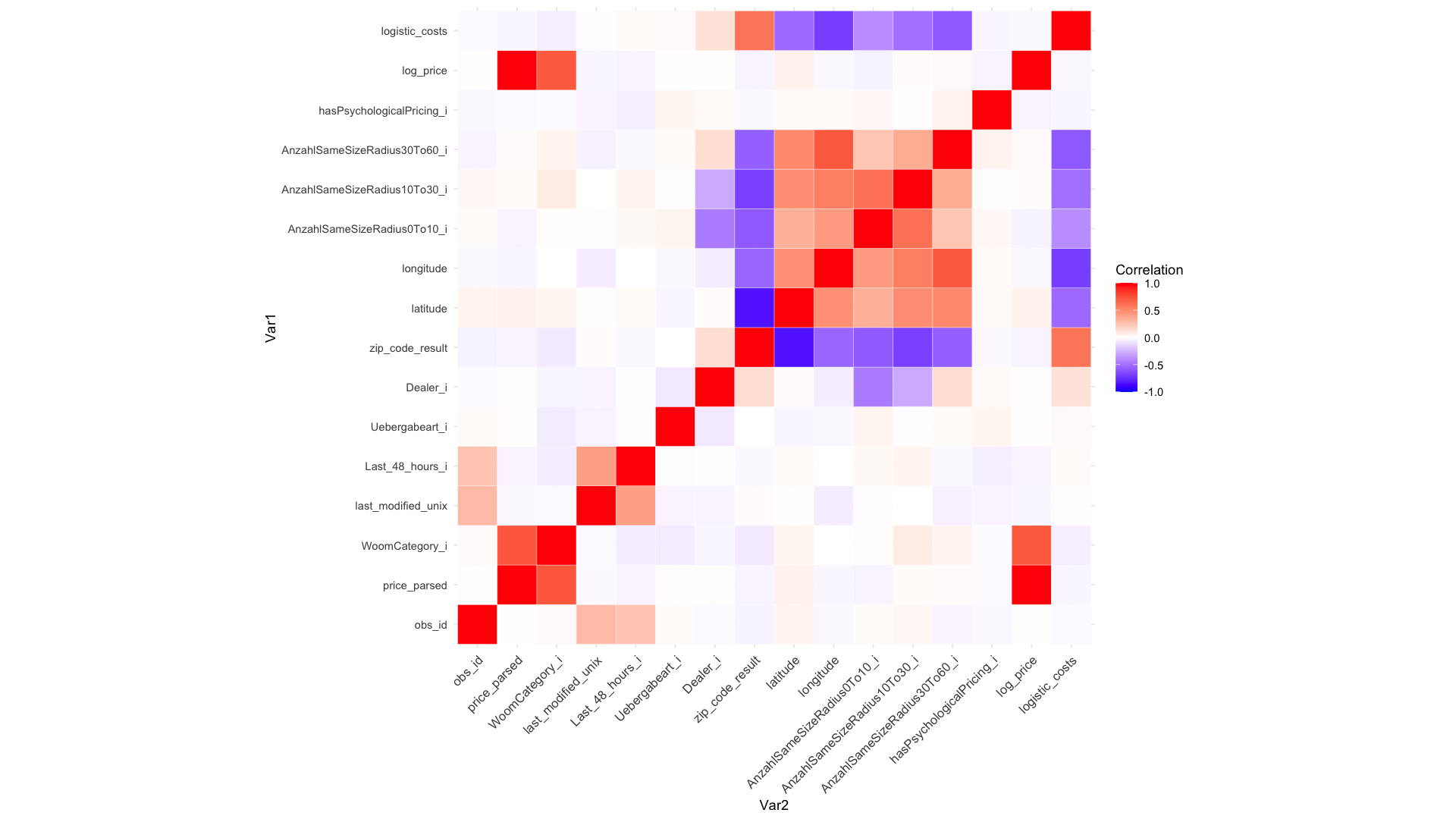}
  \caption{Correlation heatmap for all numeric variables.\newline
           \footnotesize \textit{Note:} Darker color shades indicate stronger correlations; white represents weak or no correlation. Two snapshots; \(n = 826\).}
  \label{fig:heatmap}
\end{figure}
\end{landscape}

To provide an overview of variable relationships, Figure~\ref{fig:heatmap} presents a heatmap covering all numerical variables, the categorical variable (color), and the ordinal variables (size and condition). No unexpectedly strong correlations ($\rho > $ 0.7) appear, aside from some naturally induced relationships such as those involving re-coded variables.

\subsubsection{Non-Imputed Computed Variables}

One key numerical regressor \emph{logistic\_costs} remains non-imputed whenever its value cannot be computed. For listings with no comparable bikes within a 60 km radius, the variable is set to~0. These zeroes do not represent “free” logistics but missing information; they therefore introduce measurement error, biasing the corresponding coefficient toward zero and inflating standard errors (see Table~\ref{tab:mlr_model}). The distortion is likely greatest in sparsely populated areas, so urban–rural price differentials where logistic costs should be highest may be understated by this approach.

\[
\text{logistic\_costs} \;:=\;
\begin{cases}
0, & \text{if } \text{\#count}=0, \\[6pt]
\dfrac{1}{w_\text{0-60}}, & \text{otherwise},
\end{cases}
\]

\[
\text{\#count}=N_{0\text{–}10}+N_{10\text{–}30}+N_{30\text{–}60}, 
\qquad
w_\text{0-60}= \frac{N_{0\text{–}10}}{10}+\frac{N_{10\text{–}30}}{30}+\frac{N_{30\text{–}60}}{60},
\]

\noindent where \(N_{a\text{–}b}\) is the number of comparable products located within \(a\)–\(b\) km of the listing.

A feasible imputation—such as replacing zeroes with a distance-based prediction—would mitigate this bias and produce more reliable estimates.

\subsection{Empirical Approach}

\subsubsection{Bivariate Simple Linear Regression Models}

No bivariate simple linear regression (SLR) model reached statistical significance in predicting the Willhaben ad listing price (\textit{Price}) of the bike.

\subsubsection{Stepwise AIC-Best Multiple Linear Regression Model}

Using the \textit{MASS} package in \textsf{R}, the \textit{stepAIC} function was used to run an algorithm that yields a multiple linear regression (MLR) model with the lowest (best) Akaike Information Criterion (AIC). The resulting model is presented in Table~\ref{tab:stepwise_model}. The ordinary least-squares (OLS) coefficients for the intercept and categorical variables (\textit{Size}, \textit{Condition}, and \textit{Color}) are highly statistically significant at the 1~percent level. However, the regression coefficient for the dummy variable \textit{Dealer} remains insignificant. Table~\ref{tab:stepwise_model} also displays the $F$-statistic and various other statistics. This model also reached the highest \( R^{2} \) of approximately~0.71.

\begin{landscape}
\begin{figure}[htbp]
  \centering
  \includegraphics[width=1.0\linewidth]{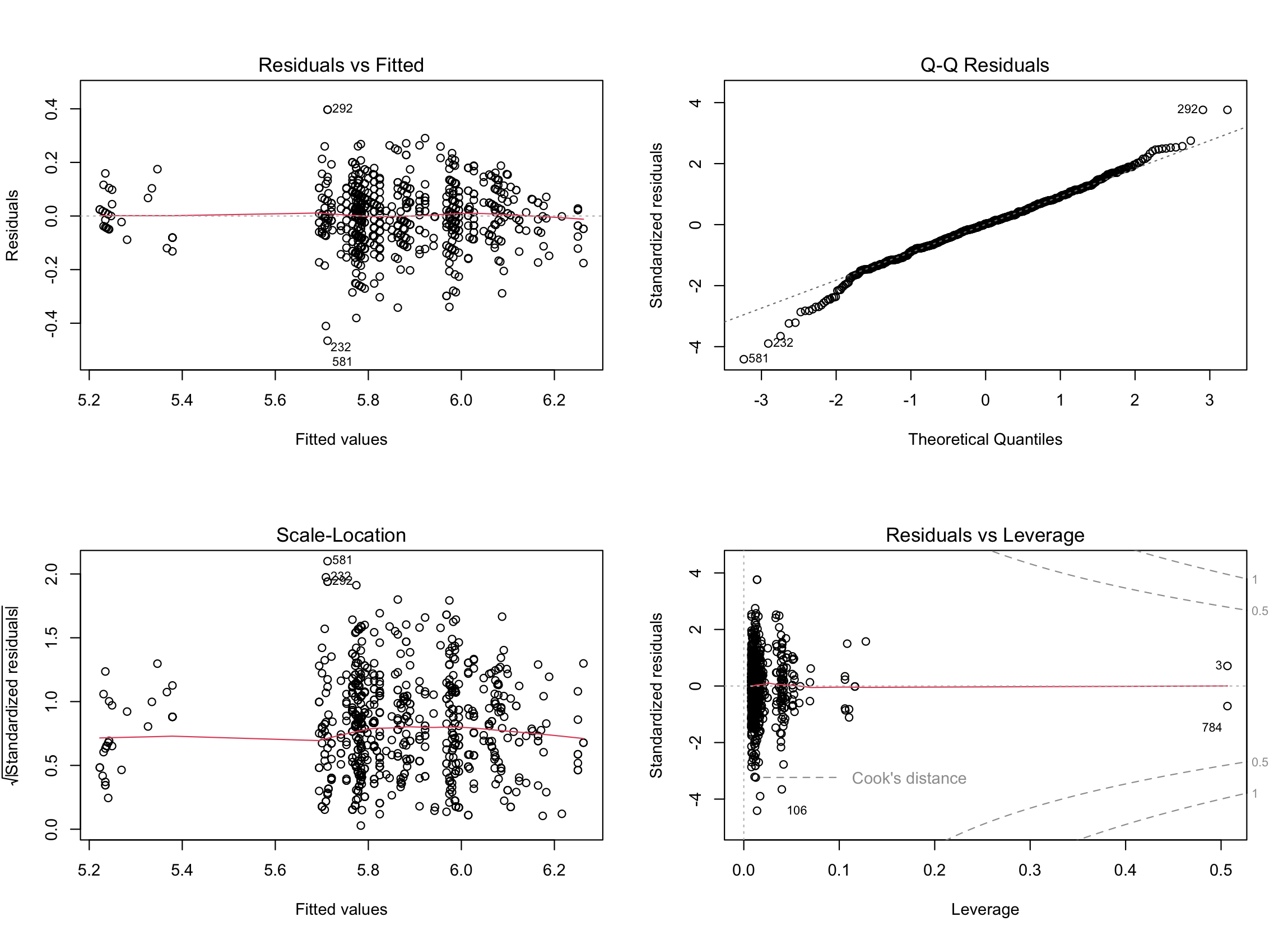}
  \caption{Regression diagnostics: Q–Q plot of regression residuals.\newline
           \footnotesize \textit{Note:} Deviations from the 45° line suggest departures from the normality assumption required for exact finite-sample inference. Two snapshots; \(n = 826\).}
  \label{fig:qqplot}
\end{figure}
\end{landscape}

\textbf{Residual analysis}: Figure~\ref{fig:qqplot} presents the residual analysis graphically, including plots of residuals vs.~fitted values, a Q–Q plot of residuals, a scale–location plot, and residuals vs.~leverage.

\subsubsection{Multiple Linear Regression Models}

Table~\ref{tab:initmodel} shows the initial run of MLR models based on a single snapshot (\( n = 254 \)) of ads. One model includes the calculated variable \textit{Logistic Costs}, defined as being inversely proportional to the sum of similar products weighted by distance. This variable might be highly biased, as it lacks an imputation strategy; however, it achieved statistical significance in the single snapshot. Including this additional variable in the regression model also slightly increased the proportion of explained variance (adjusted \( R^{2} \)).

\subsubsection{MLR Model with Fixed Effects (FE) for Size, Condition, and Color}

Table~\ref{tab:fem} presents a fixed-effects model on statistically significant categorical variables (\textit{Size}, \textit{Color}, \textit{Condition}) for the variable \textit{hasPsychologicalPricing}, which is defined as~1 if the Willhaben ad listing price contains a~9, 90, or~99. The OLS regression coefficient for \textit{hasPsychologicalPricing} is statistically insignificant at the 10~percent level but has the expected sign, as the literature suggests that such pricing strategies lead to increased consumer spending. According to Table~\ref{tab:initmodel}, the dummy variable for \textit{hasPsychologicalPricing} shifts the predicted prices upward, \textit{ceteris paribus}, by \textbf{approximately 3.4~\%}.
\subsection{Moran's \(I\) Test for Spatial Autocorrelation on Psychological Pricing (9, 90, 99 Prices)}

To test for spatial autocorrelation of the binary dummy \emph{hasPsychologicalPricing}, Table~\ref{tab:MoranPsychPricing} reports the results of Moran's \(I\) test.

\noindent The definition of Moran's \(I\), as in \citet{Moran1950}, is
\[
I=\frac{N\sum_{i=1}^{N}\sum_{j=1}^{N}w_{ij}\,(x_i-\bar{x})(x_j-\bar{x})}
        {\bigl(\sum_{i=1}^{N}\sum_{j=1}^{N}w_{ij}\bigr)\,\sum_{i=1}^{N}(x_i-\bar{x})^{2}},
\]
where \(w_{ij}\) are the spatial weights.

The associated test statistic is
\[
z(I)=\frac{I-\mathbb{E}(I)}{\sqrt{\operatorname{Var}(I)}},
\]
with \(\mathbb{E}(I)\) and \(\operatorname{Var}(I)\) denoting, respectively, the expected value and the variance of \(I\) under the null hypothesis.

\noindent Test hypotheses:
\[
H_{0}:\text{ no spatial autocorrelation }(I=\mathbb{E}(I)),\qquad 
\]
\[
H_{1}:\text{ spatial autocorrelation present }(I\neq\mathbb{E}(I)).
\]

\noindent \citet{Cliff1970} provide a historical overview of spatial autocorrelation measures, rooted in economic geography and are related to Tobler’s first law of geography.
The Moran's \(I\) test based on \(k\)-nearest neighbours (\(k=1\)) is statistically significant at the 10 \% level, yielding a very small positive value. (see Table \ref{tab:MoranPsychPricing}) This indicates that, on average, observations fall marginally in the high–high quadrant of a Moran scatterplot. Given that only 120 of 826 listings take a value of one for \emph{hasPsychologicalPricing}, this statistically weak significant spatial autocorrelation is not surprising and may simply reflect urban clusters (see, e.g., Vienna). At the conventional 5 \% level, the null hypothesis cannot be rejected, so no strong evidence of spatial autocorrelation appears for the psychological-pricing dummy. Ideally, Moran’s \(I\) should be computed within comparable groups—say, identical or neighbouring sizes (assuming buyers do not strongly prefer particular colours or conditions)—which would require a larger sample size for robust inference.

\begin{landscape}
\begin{figure}[htbp]
  \centering
  \includegraphics[width=1.1\linewidth]{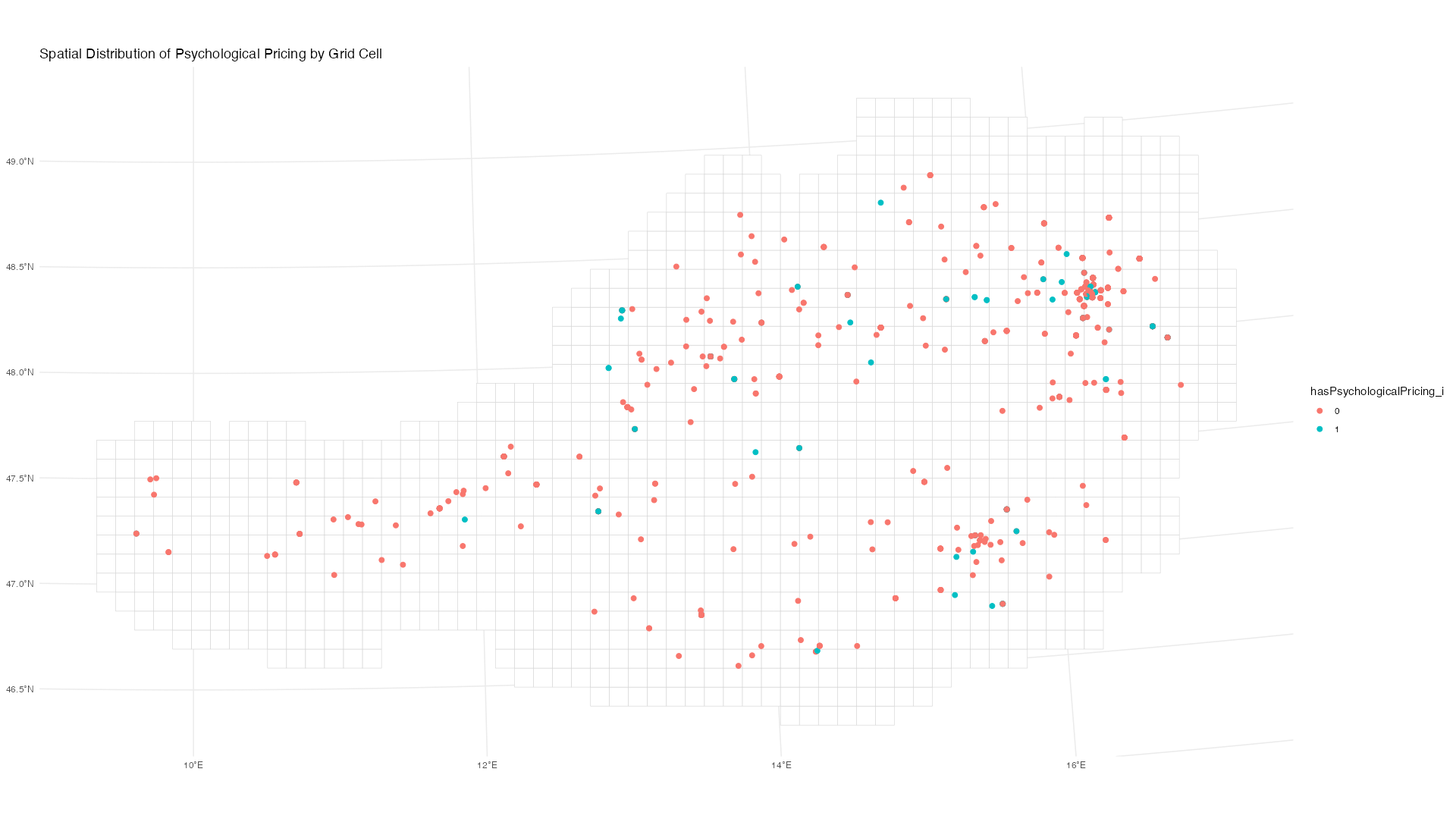}
  \caption{Spatial distribution of the psychological-pricing dummy.\newline
           \footnotesize \textit{Note:} Points are mapped onto an Austrian shapefile; colours denote presence or absence of psychological pricing. Two snapshots; \(n = 826\).}
  \label{fig:spatial_distribution}
\end{figure}
\end{landscape}

\begin{landscape}
\begin{figure}[htbp]
  \centering
  \includegraphics[width=1.1\linewidth]{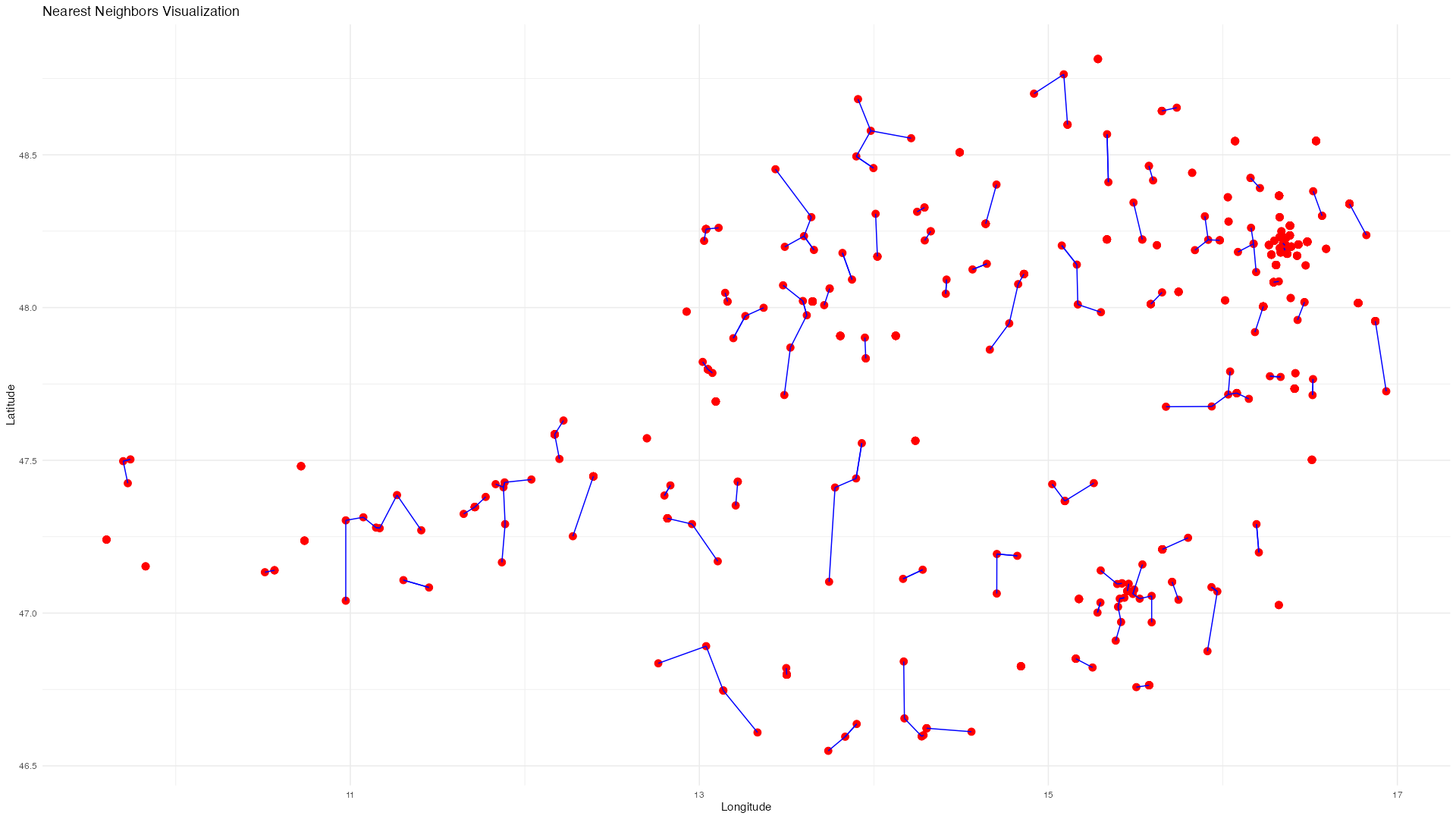}
  \caption{K Nearest-neighbour links as used for spatial weight matrix construction (\(K = 1\)).\newline
           \footnotesize \textit{Note:} Red dots mark listing locations; blue segments connect each point to its nearest neighbour, illustrating spatial proximity. Two snapshots; \(n = 826\).}
  \label{fig:nearest_neighbors}
\end{figure}
\end{landscape}

Figure~\ref{fig:spatial_distribution} maps listings with \emph{hasPsychologicalPricing} using different colour schemes. Figure~\ref{fig:nearest_neighbors} illustrates the construction of a spatial weight matrix via \(k\)-nearest neighbours (\(k=1\)) for exploring potential spatial competition hypotheses; however, this approach may be unrealistic if buyers have color preferences.

\newpage
\section{Discussion and Conclusion}

According to Hu et al.\ (2023), information asymmetry can affect market efficiency on platforms such as \textit{willhaben}. In a used-goods market the seller possesses more information about an item—its quality, signs of wear, condition, or prior use—than the buyer, creating uncertainty. Research by Moriuchi and Takahashi (2022) shows that trust therefore plays a crucial role and can lead to higher satisfaction and customer loyalty on C2C marketplaces; \textit{willhaben} addresses trust issues with its “Trusted Seller” function. If buyers perceive severe asymmetry they are less likely to purchase, yet Hu et al.\ (2023) note that when sellers outnumber buyers, disclosing too much information may hurt sales because customers find comparisons harder to make. Carefully curated information can instead differentiate listings, so \textit{willhaben} must provide—or enable—the right depth of information for both buyers and sellers.
Our empirical models identify several significant predictors of \textit{willhaben} listing prices for one product (Tables~\ref{tab:stepwise_model}), explaining much of the variation. The coefficient on \emph{logistic costs} lost significance in the two-snapshot dataset compared with the single snapshot; imputing missing values could therefore improve robustness (Tables~\ref{tab:initmodel}). Scraped data also lacked PayLivery availability, a newer feature that would equate the logistics cost of pickup with that of shipping.
Psychological-pricing strategies are increasingly relevant in e-commerce (see \citet{Hackl2014}). We found that such strategies were associated with an approx.\ 3.4\% ceteris-paribus price increase for our standardised good. Using a psychological-pricing dummy as a proxy for spatial competition, a global Moran’s \(I\) test of spatial autocorrelation found only weak statistical significance when listing ads were analysed unconditionally (Table~\ref{tab:MoranPsychPricing}); a placebo Moran’s \(I\) on logged prices showed none (Table~\ref{tab:MoranLogPrice}), suggesting product differentiation or mostly local spatial competition. A more specific hypothesis could be that if two sellers of the same Woom Bike size are geographically close and one uses a psychological price, the nearest neighbour is also more likely to do so; we found this to be true at a weak significance level by conducting a Moran's \(I\) test under randomisation of the psychological-pricing dummy without controlling for underlying product or spatial differences. For debates on whether psychological pricing is a myth using various wholesale brand items, see \citet{Ortega2023}. A spatial regression—such as a Spatial Durbin Model—should control for underlying spatial heterogeneity and, ideally, incorporate spatiotemporal data (such as the location history of official bike resellers or regional income data which we did not collect) and ad listing data collected at regular intervals, although private delistings make quantities hard to observe.
A key weakness of our study is the absence of precise data on the final transaction price and on whether the product actually sold. This gap limits our ability to analyse market behaviour and pricing strategy fully. Furthermore information asymmetry caused by the sellers or platform in a buyer's market can create pricing inefficiencies, causing buyers either to overpay or to miss favourable deals owing to unreliable information on the platform.






\newpage
\fontsize{10pt}{10pt}\selectfont


\newpage

\thispagestyle{empty}
\begin{center}
    \Large\textbf{Contributional Acknowledgements}
\end{center}

\vspace{3cm}

\noindent
\textbf{Introduction:} \\
Main contributor: Magdalena Schindl, Editor: Felix Reichel
\vspace{1cm}

\noindent
\textbf{Theoretical Part:} \\
Main contributor: Magdalena Schindl, Editor: Felix Reichel
\vspace{1cm}

\noindent
\textbf{Empirical Part:} \\
Main contributor: Felix Reichel, Editor: Magdalena Schindl
\vspace{1cm}

\noindent
\textbf{Discussion and Conclusion:} \\
Main contributor: Felix Reichel, Editor: Magdalena Schindl
\vspace{1cm}

\noindent
\textbf{Appendix:} \\
Main contributor: Felix Reichel
\vspace{1cm}

\noindent
\textbf{Submission of Supplementary Materials:} \\ Felix Reichel
\vspace{1cm}

\newpage

\appendix
\section{Supplementary Materials}
\subsection*{Tables and Figures}
    \subsection{Single snapshot MLR models}
\begin{landscape}
\begin{table}[!htbp]
\centering
\footnotesize
\caption{Multiple Linear Regression Models for Listing Prices on \textit{willhaben}\;(snapshot \(n = 254\))}
\label{tab:initmodel}
\begin{tabular*}{\textwidth}{@{\extracolsep{\fill}}lccc}
\toprule
 & \multicolumn{3}{c}{\textbf{Dependent variable: Listing price}} \\
\cmidrule(lr){2-4}
 & \(\log(\text{Price}_{i})\) & \(\text{Price}_{i}\,(\euro)\) & \(\text{Price}_{i}\,(\euro)\) \\
\midrule
\textbf{Size category} & & & \\
\hspace{1em}2 & \(0.463^{***}\) & \(122.7^{***}\) & \(122.6^{***}\) \\
\hspace{1em}3 & \(0.540^{***}\) & \(147.1^{***}\) & \(148.6^{***}\) \\
\hspace{1em}4 & \(0.739^{***}\) & \(222.4^{***}\) & \(223.8^{***}\) \\
\hspace{1em}5 & \(0.881^{***}\) & \(284.0^{***}\) & \(285.9^{***}\) \\
\hspace{1em}6 & \(0.816^{***}\) & \(250.6^{***}\) & \(252.0^{***}\) \\
\hspace{1em}7 & \(-0.024\)      & \(-3.93\)        & \(-3.93\)       \\
\hspace{1em}8 & \(0.491^{***}\) & \(134.1^{***}\) & \(133.1^{***}\) \\[2pt]
\textbf{Condition} & & & \\
\hspace{1em}good & \(0.102^{*}\)  & \(36.06^{***}\) & \(36.68^{***}\) \\
\hspace{1em}used & \(-0.080^{***}\)& \(-26.98^{***}\)& \(-26.15^{***}\) \\[2pt]
\textbf{Controls} & & & \\
Dealer\(_i\)               & \(-0.018\)        & \(-22.67^{*}\) & \(-22.67^{*}\) \\
Last 48 h\(_i\)            & \(-0.045^{*}\)    & \(-15.37^{*}\) & \(-17.01^{**}\) \\
Psychological pricing\(_i\)& \(0.034^{*}\)     & \(13.65^{*}\)  & \(13.65^{*}\)  \\
Logistic costs\(_i\)       & --                & --             & \(0.651^{*}\)  \\[2pt]
Constant                   & \(5.335^{***}\)   & \(210.2^{***}\)& \(206.3^{***}\) \\[4pt]
\textbf{Fit statistics} & & & \\
\textit{Observations}      & 254 & 254 & 254 \\
\(R^{2}\)                  & 0.646 & 0.647 & 0.651 \\
\(\bar{R}^{2}\)            & 0.628 & 0.628 & 0.632 \\
\bottomrule
\end{tabular*}

\vspace{0.5em}
\begin{minipage}{\textwidth}
\footnotesize
\textit{Notes}: Significance:\;${}^{*}p\le0.10$, ${}^{**}p\le0.05$, ${}^{***}p\le0.01$.  
Sample: \(n = 254\) of approximately 1,500 ads; snapshot scraped from \url{https://www.willhaben.at/iad/kaufen-und-verkaufen/marktplatz/fahrraeder/kinderfahrraeder-4558?keyword=woom}.  
\textbf{Logistic-costs variable}:  See section under "Non-imputed computed variables".
\end{minipage}
\end{table}
\end{landscape}

    \newpage
    \subsection{AIC-Best MLR Model}

\begin{table}[!htbp]
\centering
\caption{Stepwise AIC–best OLS regression for \(\log(\text{price})\) (two snapshots, \(n = 826\)).}
\label{tab:stepwise_model}
\footnotesize
\begin{adjustbox}{max width=0.65\textwidth}
\begin{threeparttable}
\begin{tabular}{@{\extracolsep{\fill}}lc}
\toprule
\textbf{Regressor} & \(\boldsymbol{\log(\text{price})}\) \\
\midrule
\textbf{Size category} & \\[2pt]
\hspace{1em}2 & \(0.472^{***}\)\,(0.018) \\
\hspace{1em}3 & \(0.543^{***}\)\,(0.016) \\
\hspace{1em}4 & \(0.745^{***}\)\,(0.017) \\
\hspace{1em}5 & \(0.850^{***}\)\,(0.027) \\
\hspace{1em}6 & \(0.837^{***}\)\,(0.020) \\
\hspace{1em}7 & \(-0.056^{*}\)\,(0.029) \\[4pt]

\textbf{Condition} & \\[2pt]
\hspace{1em}good & \(0.142^{***}\)\,(0.030) \\
\hspace{1em}used & \(-0.098^{***}\)\,(0.009) \\[4pt]

\textbf{Colour} & \\[2pt]
\hspace{1em}blue   & \(-0.080^{***}\)\,(0.026) \\
\hspace{1em}yellow & \(-0.072^{***}\)\,(0.027) \\
\hspace{1em}green  & \(-0.065^{**}\)\,(0.026) \\
\hspace{1em}orange & \(-0.161^{***}\)\,(0.045) \\
\hspace{1em}red    & \(-0.074^{***}\)\,(0.026) \\
\hspace{1em}violet & \(-0.033\)\,(0.026) \\[4pt]

\textbf{Controls} & \\[2pt]
Dealer\(_i\) & \(0.012\)\,(0.008) \\[6pt]
Constant     & \(5.400^{***}\)\,(0.030) \\[6pt]

\textbf{Fit statistics} &\\
\textit{Observations} & 826 \\
\(R^{2}\)            & 0.718 \\
Adjusted \(R^{2}\)   & 0.713 \\
Residual SE          & 0.106 \;(df \(= 810\)) \\
\(F\)-statistic      & 137.70\(^{***}\) \;(df \(= 15, 810\)) \\
\bottomrule
\end{tabular}
\begin{tablenotes}[flushleft]\footnotesize
\item Robust (Huber–White) standard errors in parentheses.
\item Significance levels: \({}^{*}p\le0.10\); \({}^{**}p\le0.05\); \({}^{***}p\le0.01\).
\end{tablenotes}
\end{threeparttable}
\end{adjustbox}
\end{table}

    \newpage
    \subsection{MLR Model with \# competing same size ads within radius}

\begin{table}[!htbp]
\centering
\caption{Multiple linear regression of listing prices in euro
(two snapshots, \(n = 826\)).}
\label{tab:mlr_model}
\footnotesize
\begin{adjustbox}{max width=0.8\textwidth}
\begin{threeparttable}
\begin{tabular}{@{\extracolsep{\fill}}lc}
\toprule
\textbf{Regressor} & \(\text{Price}_{i}\,(\euro)\) \\ 
\midrule
\textbf{Size category} & \\[2pt]
\hspace{1em}2 & \(121.17^{***}\)\,(4.81) \\ 
\hspace{1em}3 & \(144.21^{***}\)\,(3.77) \\ 
\hspace{1em}4 & \(219.94^{***}\)\,(4.68) \\ 
\hspace{1em}5 & \(277.96^{***}\)\,(8.13) \\ 
\hspace{1em}6 & \(261.30^{***}\)\,(8.16) \\ 
\hspace{1em}7 & \hphantom{$-$}5.25\;(5.04) \\[4pt]

\textbf{Condition} & \\[2pt]
\hspace{1em}good & \(53.94^{***}\)\,(16.34) \\ 
\hspace{1em}used & \(-35.41^{***}\)\,(3.41) \\[4pt]

\textbf{Controls} & \\[2pt]
Dealer\(_i\)                     & \hphantom{$-$}5.07\;(3.41) \\ 
Last 48 h\(_i\)                  & \(-0.34\)\;(5.07) \\ 
Psychological pricing\(_i\)      & \hphantom{$-$}2.87\;(4.62) \\[4pt]

\textbf{Same-size \#Listings (radius)} & \\[2pt]
\hspace{1em}0–10 km               & \hphantom{$-$}0.044\;(0.044) \\ 
\hspace{1em}10–30 km              & \(-0.027\)\;(0.025) \\ 
\hspace{1em}30–60 km              & \(-0.001\)\;(0.015) \\[6pt]

Constant                          & \(217.38^{***}\)\,(4.86) \\[6pt]

\textbf{Fit statistics} & \\
\textit{Observations}             & 826 \\
\(R^{2}\)                         & 0.680 \\
Adjusted \(R^{2}\)                & 0.675 \\
\bottomrule
\end{tabular}
\begin{tablenotes}[flushleft]\footnotesize
\item Robust (Huber–White) standard errors in parentheses.
\item Significance levels: \({}^{*}p\le0.10\);
      \({}^{**}p\le0.05\);
      \({}^{***}p\le0.01\).
\end{tablenotes}
\end{threeparttable}
\end{adjustbox}
\end{table}

    \newpage
    \subsection{Robustness MLR Model}

\begin{table}[!htbp]
\centering
\caption{Multiple linear regression of listing prices in euro
(two snapshots, \(n = 826\)).}
\label{tab:mlr_model_2}
\footnotesize
\begin{adjustbox}{max width=0.8\textwidth}
\begin{threeparttable}
\begin{tabular}{@{\extracolsep{\fill}}lc}
\toprule
\textbf{Regressor} & \(\text{Price}_{i}\,(\euro)\) \\ 
\midrule
\textbf{Size category} & \\[2pt]
\hspace{1em}2 & \(120.53^{***}\)\,(4.88) \\ 
\hspace{1em}3 & \(143.70^{***}\)\,(3.81) \\ 
\hspace{1em}4 & \(219.20^{***}\)\,(4.70) \\ 
\hspace{1em}5 & \(276.56^{***}\)\,(8.11) \\ 
\hspace{1em}6 & \(260.24^{***}\)\,(8.20) \\ 
\hspace{1em}7 & \hphantom{$-$}4.94\;(4.78) \\[4pt]

\textbf{Condition} & \\[2pt]
\hspace{1em}good & \(54.78^{***}\)\,(16.18) \\ 
\hspace{1em}used & \(-35.24^{***}\)\,(3.41) \\[4pt]

\textbf{Controls} & \\[2pt]
Dealer\(_i\)                 & \(4.54^{*}\)\,(2.74) \\ 
Last 48 h\(_i\)              & \(-0.50\)\,(5.09) \\ 
Psychological pricing\(_i\)  & \(2.90\)\,(4.63) \\ 
Logistic costs\(_i\)         & \(-0.50\)\,(1.95) \\[6pt]

Constant                     & \(217.43^{***}\)\,(4.51) \\[6pt]

\textbf{Fit statistics} & \\
\textit{Observations}        & 826 \\
\(R^{2}\)                    & 0.679 \\
Adjusted \(R^{2}\)           & 0.675 \\
\bottomrule
\end{tabular}
\begin{tablenotes}[flushleft]\footnotesize
\item Robust (Huber–White) standard errors in parentheses.
\item Significance levels: \({}^{*}p\le0.10\);
      \({}^{**}p\le0.05\);
      \({}^{***}p\le0.01\).
\end{tablenotes}
\end{threeparttable}
\end{adjustbox}
\end{table}

    \newpage
    \subsection{Psychological Pricing Dummy FE Model}

\begin{table}[!htbp]
\centering
\caption{Semi–log (log-level) fixed-effects model for \emph{hasPsychologicalPricing}\(_i\)
(two snapshots, \(n = 826\)).}
\label{tab:fem}
\footnotesize
\begin{threeparttable}
\begin{tabular}{@{\extracolsep{\fill}}lc}
\toprule
\textbf{Regressor} & \(\log(\text{price})\) \\ 
\midrule
hasPsychologicalPricing\(_i\) & 3.841\,(0.115) \\[6pt]
\multicolumn{2}{l}{\textit{Fixed effects included}} \\[2pt]
\hspace{1em}Size        & \checkmark \\ 
\hspace{1em}Colour      & \checkmark \\ 
\hspace{1em}Condition   & \checkmark \\[6pt]
Observations            & 826 \\
RMSE                    & 37.3 \\
Adjusted \(R^{2}\)      & 0.689 \\
\bottomrule
\end{tabular}
\begin{tablenotes}[flushleft]\footnotesize
\item Robust \(p\)-values in parentheses. Significance thresholds:
      \({}^{*}p<0.05\); \({}^{**}p<0.01\); \({}^{***}p<0.001\).
\end{tablenotes}
\end{threeparttable}
\end{table}

    \newpage
    \subsection{Moran's I Test for Psychological Pricing and Placebo Moran's I Test}

\begin{table}[!htbp]
\centering
\caption{Moran’s \(I\) test (randomisation) for the psychological-pricing dummy; two snapshots, \(n = 826\).}
\label{tab:MoranPsychPricing}
\footnotesize
\begin{threeparttable}
\begin{tabular}{lS}
\toprule
Statistic & {Value} \\
\midrule
Moran \(I\) standard deviate & 1.46 \\
\(p\)-value                 & \textbf{0.07} \\
Expectation                 & -0.0012 \\
Variance                    & 0.0015 \\
\bottomrule
\end{tabular}
\end{threeparttable}
\end{table}

\begin{table}[!htbp]
\centering
\caption{Placebo Moran’s \(I\) test (randomisation) for \(\log(\text{price})\); two snapshots, \(n = 826\).}
\label{tab:MoranLogPrice}
\footnotesize
\begin{threeparttable}
\begin{tabular}{lS}
\toprule
Statistic & {Value} \\
\midrule
Moran \(I\) standard deviate & -0.63 \\
\(p\)-value                 & 0.73 \\
Expectation                 & -0.0012 \\
Variance                    & 0.0015 \\
\bottomrule
\end{tabular}
\end{threeparttable}
\end{table}

    \newpage



\subsection{Figures}


    \newpage
    
\end{document}